\documentclass[aip,pop,reprint,superscriptaddress]{revtex4-1}
\usepackage{amsmath}
\usepackage{graphicx}% Include figure files
\usepackage{dcolumn}% Align table columns on decimal point
\usepackage{bm}% bold math
\usepackage{float}
\begin{document}

%\preprint{APS/123-QED}
%\begin{CJK*}{GB}{gbsn}  %%%%\begin{CJK*}{GBK}{song}
\title{Study of filamentation instability on the divergence of ultraintense laser-driven electrons}

% Force line breaks with \\

\author{X. H. Yang}\thanks{Electronic mail: xhyang@nudt.edu.cn}
\affiliation{College of Science, National University of Defense Technology, Changsha 410073, China}

\author{H. B. Zhuo}\affiliation{College of Science, National University of Defense Technology, Changsha 410073, China}

\author{H. Xu}\affiliation{School of Computer Science, National University of Defense Technology, Changsha 410073, China}

\author{Z. Y. Ge}\affiliation{College of Science, National University of Defense Technology, Changsha 410073, China}

\author{F. Q. Shao}\affiliation{College of Science, National University of Defense Technology, Changsha 410073, China}

\author{M. Borghesi}\affiliation{School of Mathematics and Physics, Queen's University of Belfast, Belfast BT7 1NN, United Kingdom} \affiliation{Institute of Physics of the ASCR, ELI-Beamlines Project, Na Slovance 2, 18221 Prague, Czech Republic}

%\author{W. Yu}
%\affiliation{Shanghai Institute of Optics and Fine Mechanics, Chinese Academy of Sciences, Shanghai 201800, China}

\author{Y. Y. Ma}\thanks{Electronic mail: yanyunma@126.com}
\affiliation{College of Science, National University of Defense Technology, Changsha 410073, China}


\date{\today}% It is always \today, today, but any date may be explicitly specified

\begin{abstract}
Generation of relativistic electron (RE) beams during ultraintense laser pulse interaction with plasma targets is studied by collisional particle-in-cell (PIC) simulations. Strong magnetic field with transverse scale length of several local plasma skin depths, associated with RE currents propagation in the target, is generated by filamentation instability (FI) in collisional plasmas, inducing a great enhancement of the divergence of REs compared to that of collisionless cases. Such effect is increased with laser intensity and target charge state, suggesting that the RE divergence might be improved by using low-Z materials under appropriate laser intensities in future fast ignition experiments and in other applications of laser-driven electron beams.
\end{abstract}

\pacs{52.38.Kd, 52.35.Qz, 52.65.Rr}
% PACS, the Physics and Astronomy Classification Scheme. %%

\maketitle
%\end{CJK*}

\section{Introduction}\label{sec:1}
Ultraintense laser-driven REs have attracted great recent attention \cite{Yang15,Zhuo14} due to their potential application in the areas of fast ignition laser fusion \cite{Tabak94,Yang12,Shiraga14,Robinson14}, ion acceleration by laser-plasma interaction \cite{Macchi13,Yang10}, and production of ultrashort bright radiations \cite{Jin11}. It is essential to characterize accurately the RE divergence for these applications.

Both the cause and characterization of the RE divergence are still not understood well, which are
critically important to determine the RE energy deposition in targets.
The divergence measured in the experiments usually increases with laser intensity \cite{Green08}, though some discrepancy exists among the different diagnostic techniques because each one is dependent on different parameters \cite{Lancaster07}. However, recent PIC simulations show that the RE divergence is approximately linearly proportional to the preplasma scale length for a fixed laser intensity, but is weakly dependent on the laser intensity for a fixed preplasma \cite{Ovchinnikov13}. For a normally incident laser, the electron injection angle $\theta_i=\tan^{-1}\sqrt{2/(\gamma-1)}$ is obtained from the electron's trajectory in an intense electromagnetic wave \cite{Moore95}, where $\gamma$ is the electron relativistic factor. However, the scaling predicts a decrease of RE divergence as the laser intensity increases, which disaccords with both experimental and numerical results. Adam et al. \cite{Adam06} found that the RE divergence is mainly induced by the large static magnetic fields generated in the laser-plasma interaction layer, arisen from the Weibel instability driven by a thermal anisotropy of the electrons. Furthermore, the full divergence of REs including a regular radial beam deviation and a random angular dispersion is reported recently \cite{Debayle10}, which are determined by the transverse component of the laser ponderomotive force and collisionless Weibel instability induced micro-magnetic fields, respectively. The Weible/filamentation instability induced magnetic field will be saturated after a rapid linear growth due to magnetic trapping \cite{Davidson72,Yang94,Okada07}. In addition, magnetic field of $10^4$T magnitude generated by the reflected laser in preplasma can also deflect the laser-driven REs \cite{Perez13}.

It is found that transverse temperature of beams can suppress the FI under certain conditions as it propagates in cold plasmas both in theoretical analysis and numerical simulations, but for a collisional plasma, the growth rate of instability will increase with plasma collision rate \cite{Molvig75,Karmakar08} or plasma resistivity (that is proportional to the collision rate) \cite{Gremillet02}, which can only be inhibited by external magnetic field \cite{Molvig75,Kapetanakos74}.
In addition, collisional effects tend to attenuate the current FI for symmetric (where $n_h=n_b$, $n_h$ and $n_b$ are the relativistic and background electron density, respectively) or quasisymmetric counterstreaming while enhance it for extremely asymmetric counterstreaming (where $n_h\ll n_b$) \cite{Hao09}, which is in the context of laser-driven REs propagating in solid or compressed targets. Note that only the influence of magnetic field generated around the laser-plasma interaction region on the RE divergence and usually collisionless plasmas are considered in the previous studies. However, since the collision frequency of background plasma can be comparable to the local plasma frequency in solid targets, the collisions of plasmas have to be considered properly as the RE propagates in it\cite{Volpe13}. Strong magnetic fields are ubiquitous during the RE beams propagating in dense target due to growth of FI induced by plasma collisions \cite{Yang15P,Schmitz12,Heron15}. Such magnetic fields can significantly enhance or suppress the RE divergence, which is not yet fully understood and needs to be investigated in detail.

In this paper, angular distribution of laser-driven REs with laser intensity varying from $10^{18}$ W/cm$^2$ up to $5\times10^{20}$ W/cm$^2$ and target materials of copper (Cu), aluminium (Al), and polymer (CH$_{2}$) are studied by collisional PIC simulations. It is found that strong magnetic field is generated by the collisional FI during the REs propagating in collisional plasmas, inducing a great enhancement of the divergence of REs compared to that of collisionless cases. The divergence of REs increases with laser intensity and target charge state. This finding would be helpful for future fast ignition experiments and in other applications of laser-driven electron beams.

\section{Simulation model}\label{sec:2}
To investigate the FI induced magnetic field on RE divergence in collisional plasmas, numerical simulations are performed using the relativistic collisional 2D3V PIC code EPOCH \cite{Arber15}, which includes the binary collision model of charged particles proposed by Sentoku et al. \cite{Sentoku08}. The target consists of an initially neutral mixture of electrons and Cu ions with mass $m_{Cu}=63.5m_p$, where $m_p=1836m_e$ is the proton mass. The charge state and maximum density of the ions are set to 15 and $28n_c$, where $n_c=9.92\times10^{20}$cm$^{-3}$ is the critical density, corresponding to the laser wavelength $\lambda_L=1.06\mu$m. The plasma has an exponentially increasing density profile $n_e=5n_c\exp((x-10\mu m)/1\mu m)$ between $x=6-10\mu m$ and then keeps the maximum electron density of $420n_c$ between $x=10-25\mu m$. Note that $\sim33\%$ of the solid density of Cu ions is used in the simulations, in order to lower the computational costs of PIC simulations at the required resolution for full solid density. The initial temperature of the electrons and Cu$^{15+}$ is 100 eV. The simulation box is $30\mu m\times20 \mu m$ with $3000 \times 2000$ cells. Each cell contains 50 numerical macroparticles per species in the simulations. A $p$-polarized laser pulse with a peak intensity of $I_L=10^{20}$W/cm$^2$ is incident normally from the left boundary, corresponding to $a_L=eE_L/m_ec\omega_L=9.06$ for the dimensionless maximum amplitude of the laser electric field, where $e$ and $m_e$ are the electron charge and mass, $c$ is the light speed, $E_L$ and $\omega_L$ are the electric field and frequency of the laser pulse, respectively. The spatial profiles of the laser is Gaussian, with spot radius $4 \mu m$. The laser pulse rises up with a Gaussian profile over the first 15 fs to the peak intensity, stays at constant intensity for 105 fs. In order to suppress the numerical heating, fifth-order interpolation scheme is employed to evaluate the currents. Periodic and absorbing boundary conditions are used for the transverse and longitudinal boundaries, respectively.

\section{Results and discussion}\label{sec:4}
\begin{figure}\suppressfloats\centering
\includegraphics[width=8.5cm]{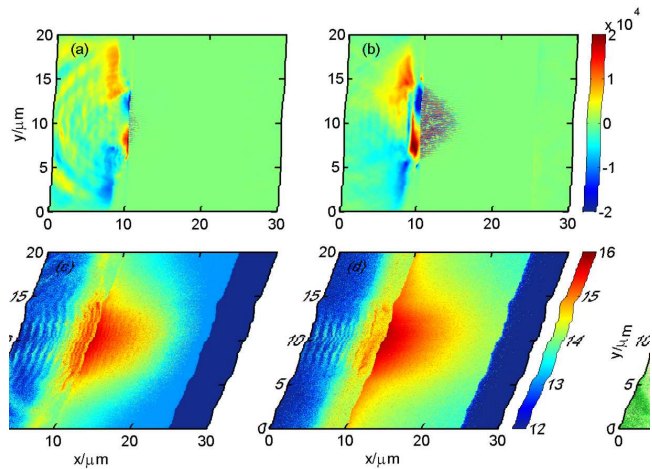}%
\caption{\label{f1} Distributions of the self-generated magnetic field ($B_z$) [(a) and (b)] and $log_{10}$ of kinetic energy density of the electrons [(c) and (d)] at t=100 fs [(a) and (c)] and 160 fs [(b) and (d)] for the case of a collisional Cu target, respectively.  The magnetic field is in units of T and the kinetic energy density is in units of $J/m^3$ (same as in the other figures).}
\end{figure}

\begin{figure}\suppressfloats\centering
\includegraphics[width=8.5cm]{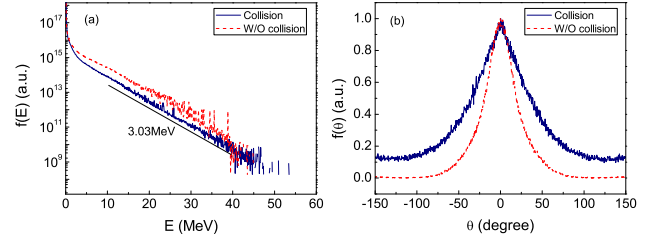}%
\caption{\label{f5} Energy spectra of the electrons for collisional and collisionless Cu targets (a). Angular distribution of REs ($E_k\geq50$ keV) for the collisional and collisionless Cu targets at t=100 fs (b), respectively, in which the electrons are extracted from the region of $x=11-24\mu m$ and $y=4-16\mu m$ (same as in Fig. \ref{f6}(c) and Fig. \ref{f7}(c)).}
\end{figure}

Figure \ref{f1} shows the self-generated magnetic field and the electron kinetic energy density distributions for the case of a collisional Cu target. It can be seen that two types of large scale magnetic fields having distinct polarities are generated in the preplasma and around the solid target interface (i.e., $x=10\mu m$). The magnetic field generated in the preplasma is mainly arisen from $\nabla n_e\times\nabla T_e$ due to the gradients of electron density and temperature, whose magnitude can be estimated as \cite{Bell93} $B\sim200(\frac{\tau}{ps})(\frac{k_BT_e}{keV})(\frac{L_T}{\mu m})^{-1}(\frac{L_n}{\mu m})^{-1}$T, where $T_e$ is the electron temperature, $\tau$ is the pulse duration, $L_T$ and $L_n$ are the (transverse) temperature and (longitudinal) density gradient, respectively. $L_T$ and $L_n$ can be estimated from the electron temperature and density distributions, which are around $5.5\mu m$ and $1.4\mu m$, respectively, and the electron temperature is $3.03$ MeV (see Fig. \ref{f5}(a)) in the preplasma. Thus the magnitude of $B$ is $9.4\times10^3$T, which is very close to the simulation results. The field around the solid target interface is driven by temporal variations in the ponderomotive force and can be estimated by $\nabla^2B\sim\nabla n_e\times\nabla I$ \cite{Gibbon05}, where $I$ is the laser intensity. In addition to the large scale magnetic fields, microscale magnetic field with periodical distribution is also observed in the dense target, whose magnitude can reach $1.5\times10^4$T. The fields is associated with the RE propagation in the target, as shown in Fig. \ref{f1}(c) and (d), in which the REs are beamed into the target and heat the target rapidly. The kinetic energy density of the electrons can be greater than $10^{16}J/m^3$.

\begin{figure}[h]\suppressfloats\centering
\includegraphics[width=8.5cm]{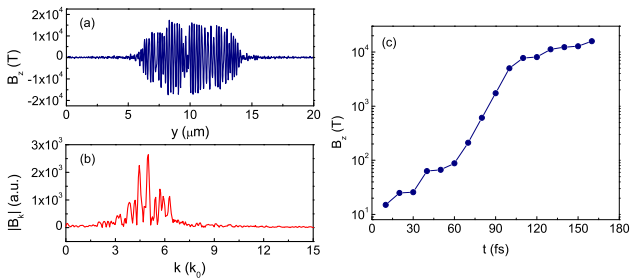}%
\caption{\label{f2} Profile of $B_z$ along the $y$ direction at $x=11\mu m$ at t= 160fs (a), which is averaged over 1$\mu m$ along the $x$ direction. The corresponding frequency spectrum $|B_k|$ of $B_z$ (b). Evolution of $B_z$ with time in the front surface of the solid target around ($x=11\mu m, y=8\mu m$) (c).}
\end{figure}

The profile of $B_z$ along $y$ direction at $x=11\mu m$ and corresponding frequency spectrum $|B_k|$ of $B_z$ are presented in Fig. \ref{f2}(a) and (b).
It is shown that peak of the wave number of the magnetic field is $\sim5k_0$, where $k_0=2\pi/1\mu m\approx k_L$, $k_L$ is the wave number of laser pulse. The local plasma skin depth $\lambda_{p}=2\pi c/\omega_{p}=0.049\lambda_L$, where $\omega_{p}=(4\pi n_ee^2/m_e)^{1/2}$ and $n_e$ are the local electron frequency and density. Thus, the transverse scale length of the magnetic field is $\sim4\lambda_p$. The growth rates of modes whose wavelength is comparable to the plasma skin depth are suppressed and the spectral peak of the growth rate shifts to long wavelength modes compared to that of collisionless case (not shown for brevity), similar to the previous theoretical and numerical results \cite{Sentoku00,Hao09}. Thus, magnetic fields with relatively large structures are observed in the collisional plasmas, as shown in Fig. \ref{f1}(b) and Fig. \ref{f2}(a).

Figure \ref{f2}(c) shows the evolution of self-generated magnetic field $B_z$ with time in the front surface of the solid target. It is seen that the magnetic field grows slowly before $t=60$fs, however, it grows rapidly with a linear growth rate as the REs propagates into the target during $t=60-100$fs. The maximum growth rate of the magnetic field is estimated to be $\Gamma_{max}=1.07\times10^{14}/s\approx0.03\omega_{p}$.
Assuming that the only role of collisions is to slow down the REs, for simplicity, the maximum growth rate of FI for a "cold" beam propagating in collisionless plasmas is employed \cite{Bret10}
\begin{equation}\label{growth}
\Gamma\sim\beta_{h}\sqrt{\frac{\kappa}{\gamma_{h}}},
\end{equation}
where $\beta_{h}=v_h/c$, $v_h$ is the RE velocity along the laser propagation axis, $c$ is the light speed in vacuum, $\gamma_{h}$ is the Lorentz factor of REs, and $\kappa=n_h/n_b$ is the beam to plasma density ratio. From the RE temperature here (3.03MeV, see Fig. \ref{f5}(a)), we get $\gamma_{h}\approx6.93$ and $\beta_{h}=0.9895$. The laser absorption efficiency here is $\alpha=0.42$, and the RE density can be obtained by conservation of energy flux
$\alpha I_L=n_hv_hT_h$, which leads to a RE density of $n_h=2.93\times10^{21}cm^{-3}$. Thus, $\kappa=0.004$ and $\Gamma=0.024\omega_{p}$, which is comparable to the simulation results. It is seen that the field becomes saturated nonlinearly with a magnitude of $\sim1.5\times10^4$T after $t=120$fs. Such nonlinear saturation appears as the gyroradius of REs becomes of the order of the modulation wavelength, i.e., $r_L\sim \lambda_F$, namely magnetic trapping\cite{Davidson72,Yang94}, leading to a saturation magnetic field amplitude
\begin{equation}\label{Bs}
B_s=\frac{\gamma_h m_ev_y}{e\lambda_F},
\end{equation}
where $v_y\approx0.4c$ is the transverse velocity of REs that is estimated from the RE temperature and divergence (i.e., $43.4^\circ$, see Fig. \ref{f5}(b)), $\lambda_F=4\lambda_p$ is applied here.
Thus, the saturated magnetic field can be estimated to be $B_s=2.25\times10^4$T, which has the same order with that observed in the PIC simulation.

\begin{figure}[h]\suppressfloats\centering
\includegraphics[width=8.5cm]{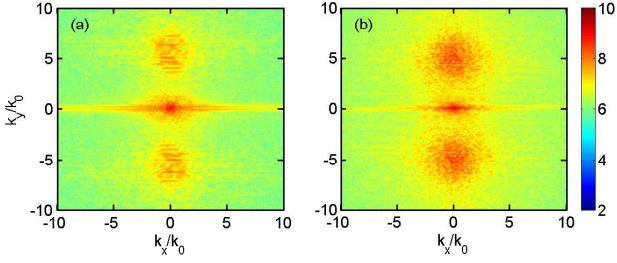}%
\caption{\label{f3} Fast Fourier transform of $B_z$ at t=100 fs (a) and 160 fs (b), respectively.}
\end{figure}

In order to understand better the filamentation instability observed here, we show the Fast Fourier transform of $B_z$ at t=100 fs and 160 fs in Fig. \ref{f3}. It can be seen that transverse modes appear with time evolution, and a clear maximum wave number of around $k_y=5k_0$ is observed, which persists that in Fig. \ref{f2}. In addition, the peak spread widely in $k_x-k_y$ space, indicating that the modes are not purely transverse but oblique. This is consistent with the theoretical results \cite{Gremillet07,Hao12} that electromagnetic oblique modes can be enhanced by collisional effects and FI is the dominant mode for a beam propagating in dense plasmas. Note that the peak around $\mathbf{k}=0$ is attributed to the fact that $\mathbf{k}$ becomes 0 as the laser pulse arrives at the critical point of the target.

\begin{figure}\suppressfloats\centering
\includegraphics[width=8.5cm]{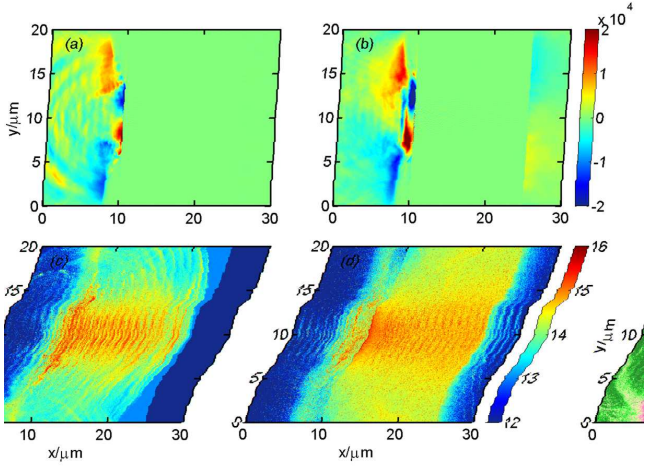}%
\caption{\label{f4} The same as that in Fig. 1, but for a collisionless Cu target.}
\end{figure}

For comparison, ultraintense laser interaction with a collisionless Cu target is also investigated, as shown in Fig. \ref{f4}. It is seen that, large scale magnetic fields having distinct polarities are generated in the preplasma and around the solid target interface, which is similar to that in the collisional case. However, periodic magnetic field is not obvious in the collisionless plasma bulk, indicating that the plasma only experiences very weak instability growth, which is consistent with the results reported in Ref. \cite{Karmakar08} that collisionless FI is completely shut down for a RE beam with a large transverse temperature. The instability stabilized by the RE transverse energy spread because of beam transverse temperature has already been mentioned in many previous literatures \cite{Molvig75,Fiore10,Silva02}.
From the RE divergence, it can be estimated that the transverse temperature of REs is 480keV and a ratio of $T_y/T_x$ about 0.16, where $T_x$ and $T_y$ are the RE temperature in the $x$ and $y$ directions, respectively, suggesting that transverse energy of the REs spreads remarkably during the RE propagating in the target. That is, the collision guarantees the occurrence of FI regardless of beam transverse temperature here.
On the other hand, in collisionless plasma, the FI would be suppressed significantly as the ratio of beam to plasma density gets very small \cite{Hao09}.
From the comparison, it is also predicted that the growth rate of the instability increases with the collision frequency in the case of RE beam and small ratio of beam to plasma.
The REs propagate mainly in the forward direction in the collisionless case, as shown in Figs. \ref{f4}(c) and (d), while they propagate forward with a parabolic profile in the collisional case (Figs. \ref{f1}(c) and (d)), indicating that the REs should have smaller divergence in the former case comparing to that of the latter.
It is noted that the RE distribution is more diffuse than that in Fig. \ref{f1}, which could be due to the absence of confinement of the FI induced magnetic fields that scatter the REs but can also confine the latter due to its very high intensity and wide spread.

Figure \ref{f5}(a) shows the energy spectra of the electrons both for the collisional and collisionless Cu targets. It is shown that the spectra have similar profiles in these cases.
The RE temperature is 3.03MeV for the collisional target and is identical with that of the collisionless case. It can be attributed to the fact that the REs here are mainly accelerated by the $\mathbf{J}\times\mathbf{B}$ heating \cite{Kruer85}, which can be clearly seen from the electron kinetic energy density distributions (Fig. \ref{f1} and Fig. \ref{f4}), where the REs are separated by a distance of half of a wavelength, so the electron temperature is weakly affected by the collision frequency. The laser absorption efficiency for the collosionless case is nearly identical to that of collisional case ($\sim$0.38). However, the number of the REs for the collisional case is a bit lower than that of the collisionless case due to the microscale magnetic fields arisen from the filamentation instability. Note that the RE temperature here is lower than that given by ponderomotive scaling \cite{Wilks92} $T_h=m_ec^2(\sqrt{1+a_{L}^2}-1)=4.14$MeV, which can be attributed to the fact that the electrons only interact with the laser pulse during a fraction of the laser cycle before being accelerated forward beyond the laser penetration region due to the relatively short preplasma scale-length employed in our simulation \cite{Haines09}.

Angular distribution of REs ($E_k\geq50$ keV) is shown in Fig. \ref{f5}(b). The electron momentum angle here is determined by $\theta=\tan^{-1}(p_y/p_x)$, where $p_y$ and $p_x$ are the electron momentum in the $y$ and $x$ directions, respectively. Since the relatively short length for target employed in the simulations, we mainly focus on the RE divergence at $t=100$fs to avoid refluxing of REs from rear side of the target. The electron angular distribution can be fitted by a Gaussian function
\begin{equation}\label{Gaussian}
f_F(\theta)=f_0+\frac{A}{\sigma \sqrt{\pi/2}}e^{-2\frac{(\theta-\theta_c)^2}{\sigma^2}}.
\end{equation}
where $f_0$ and $A$ are constant, $\theta_c$ and $\sigma$ are electron mean propagation angle and dispersion angle, which are related to the beam transverse velocity and the electron transverse temperature, respectively. It is shown that the full width at half max (FWHM) of the RE divergence reaches 66.9$^\circ$ for the collisional case, which is significantly greater than that of the collisionless case ($43.4^\circ$). This increase is attributed to the fact that the collisional FI induced magnetic field effectively scatters the REs \cite{Adam06}. In the presence of such microscale magnetic field, the mean transverse velocity of the electrons is zero, but the mean square velocity is not, which determines the mean square angular of the electrons and leads to a large electron divergence \cite{Adam06}.

\begin{figure}[h]\suppressfloats\centering
\includegraphics[width=8.5cm]{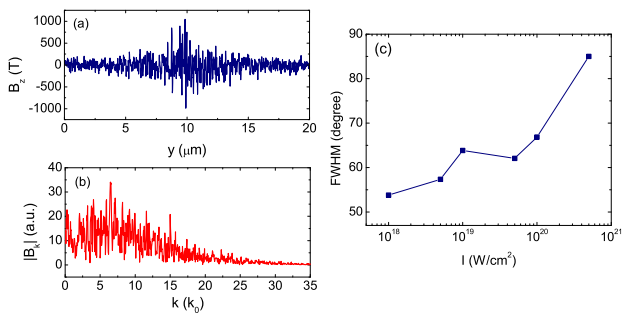}%
\caption{\label{f6} The same as that in Fig. 2, but for a laser intensity of 10$^{19}$W/cm$^2$ [(a) and (b)]. FWHM of the RE ($E_k\geq50$ keV) angular distribution as a function of laser intensity for a collisional Cu target (c).}
\end{figure}

Figure \ref{f6}(a) and (b) shows the profile of $B_z$ along $y$ direction at $x=11\mu m$ and corresponding frequency spectrum $|B_k|$ of $B_z$ for the case with a laser intensity of 10$^{19}$W/cm$^2$. It can be seen that the magnetic field here is relatively weaker with an amplitude less than 500T, and there is no obvious wave number peak in its frequency spectrum, meaning that the growth of filamentation instability is very tiny. It is due to the fact that lower RE current density is produced for the lower laser intensity compared to that of the high laser intensity. The laser absorption efficiency here is 0.48 that is slightly higher than that of the aforementioned case. The dependence of RE divergence on laser intensity is presented in Fig. \ref{f6}(c). Since the magnitude of self-generated magnetic field increases with laser intensity due to higher RE current generation, it can scatter the REs more effectively. It is shown that the divergence increases with laser intensity, which is in good agreement with the experimental results \cite{Green08}. This is different with the results \cite{Cui13} that higher laser intensity leads to smaller RE divergences, where only the divergence of much higher energy electrons is considered. It is also distinct from that in collisionless cases, in which the divergence of REs is weakly dependent on the laser intensity (not shown for brevity). The discrepancy indicates that properties of target should be considered comprehensively for investigating RE characteristics, especially for solid target, in which the collision could have a significant influence on the RE divergence.

\begin{figure}[h]\suppressfloats\centering
\includegraphics[width=8.5cm]{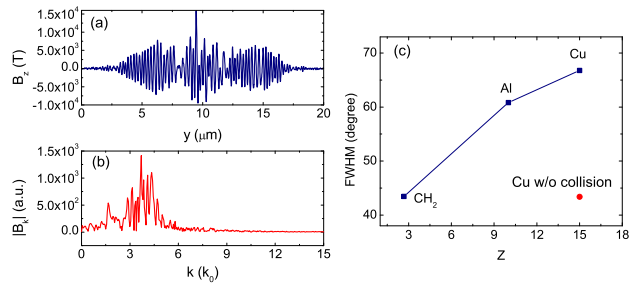}%
\caption{\label{f7} The same as that in Fig. 2, but for a collisional CH$_2$ target [(a) and (b)]. FWHM of the RE ($E_k\geq50$ keV) angular distribution for the cases of collisional Cu, Al, and CH$_2$ targets (c). The laser intensity is fixed at 10$^{20}$W/cm$^2$.}
\end{figure}

For completely, the RE divergence for cases of ultraintense laser interaction with various materials is also studied, as shown in Fig. \ref{f7}. The charge state and maximum ion density for CH${_2}$ and Al targets are set to 2.67, 10, 45$n_c$, and 25$n_c$, respectively.
In order to keep an identical target density that generates REs, similar preplasma profiles to that in Cu target are used both for CH${_2}$ and Al targets.
Note that the energy spectrum profiles of the electrons are very close to each other in these three cases (not shown for brevity), suggesting that the effect of the electron temperature differences for different targets can be neglected in this investigation.
This is different with the results reported in Ref. \cite{Chen04} that the RE electron temperature is dependent on the target materials, in which a much lower laser intensity of $10^{16}$W/cm$^{2}$ is employed, and the electrons are mainly accelerated by the resonance acceleration or vacuum heating (if the scale length of preplasma is less than the wavelength) \cite{Cui13}. In these regimes, the RE temperature is dependent on the scale length of the preplasma, which are different for different material targets because the plasma hydrodynamic expansion from lower-Z targets is faster than that of high-Z targets.
From Fig. \ref{f7}(a) and (b), it is seen that microscale magnetic field similar to that in Fig. \ref{f2} is generated in CH$_2$ target, while the magnetic field is relatively weak (5500T). The peak of the wave number is $\sim3.7k_0$, corresponding to a wave length of 5.5$\lambda_p$, slightly larger than that of Cu target. The RE divergence for CH${_2}$ target is only 43.4$^\circ$ due to the relatively weak magnetic field in the target, which is equal to that in collisionless Cu target.
Figure \ref{f7}(c) shows that the divergence increases rapidly with target charge state.
It is because of that the collision frequency of electron-ion given by \cite{Ma12} $\nu_{ei}=\frac{4\sqrt{2\pi}}{3}\frac{n_iZ_{i}^{2}e^4}{m_{e}^{1/2}T_{e}^{3/2}}\ln\Lambda$ increases with plasma density and charge state, where $n_i$ and $Z_i$ are the ion density and charge state, $\ln\Lambda$ is the Coulomb logarithm. That is, the particles would experience higher collisions in high-Z targets compared to that in low-Z targets.
Since the growth of FI in case of extremely asymmetric counterstreaming is significantly suppressed only except that plasma collisions are considered \cite{Hao09}, as observation in our simulations, stronger magnetic field generated and larger RE divergence induced in Cu target compared to Al(or CH$_2$) target.

\section{Conclusion}\label{sec:5}
In conclusion, angular distribution of laser-driven REs with laser intensity varying from $10^{18}$ W/cm$^2$ up to $5\times10^{20}$ W/cm$^2$ and target materials of Cu, Al, and CH$_2$ are studied by collisional PIC simulations. Intense microscale magnetic field is generated by the FI during the REs propagating in collisional plasmas, and such instability is almost suppressed by the transverse temperature of REs (also due to the small ratio of beam to plasma density) in collsionless target, leading to a great enhancement of the RE divergence in collisional plasmas compared to that of collisionless cases. The divergence of REs increases with laser intensity and target charge state due to intenser magnetic field generation for higher laser intensities and high-Z targets. The results should be helpful for applications of laser-driven electron beams.

\begin{acknowledgments}\suppressfloats
This work was supported by the NNSFC (11305264, 11275269, 11375265, and 91230205) and the Research Program of NUDT.
%M.B. acknowledges funding from projects ELI (Grant No. CZ.1.05/1.1.00/483/02.0061) and OPVK 3 (Grant No. CZ.1.07/2.3.00/20.0279) and from EPSRC (Grant Nos. EP/K022415/1 and EP/I029206/1).
The authors wish to acknowledge Prof. M.Y. Yu (ZJU) and Dr. H.Schmitz (STFC RAL) for fruitful discussions.
\end{acknowledgments}

\newpage


\begin{thebibliography}{99}\suppressfloats
% Note: do not put a dot after the journal name unless when the last word is an abbreviation!

\bibitem{Yang15} X. H. Yang, W. Yu, H. Xu, M. Y. Yu, Z. Y. Ge, B. B. Xu, H. B. Zhuo, Y. Y. Ma, F. Q. Shao, and M. Borghesi, Appl. Phys. Lett. {\bf 106}, 224103 (2015).

\bibitem{Zhuo14} H. B. Zhuo, Z. L. Chen, Z. M. Sheng, M. Chen, T. Yabuuchi, M. Tampo, M. Y. Yu, X. H. Yang, C. T. Zhou, K. A. Tanaka, J. Zhang, and R. Kodama, Phys. Rev. Lett. {\bf 112}, 215003 (2014).

\bibitem{Tabak94} M. Tabak, J. Hammer, M. E. Glinsky, W. L. Kruer, S. C. Wilks, J. Woodworth, E. M. Campbell, M. D. Perry, and R. J. Mason, Phys. Plasmas {\bf 1}, 1626 (1994).

\bibitem{Yang12} X. H. Yang, M. Borghesi, and A. P. L. Robinson, Phys. Plasmas {\bf 19}, 062702 (2012).

\bibitem{Shiraga14} H. Shiraga, H. Nagatomo, W. Theobald, A. A. Solodov, and M. Tabak, Nucl. Fusion {\bf 54}, 054005 (2014).

\bibitem{Robinson14} A. P. L. Robinson, D. J. Strozzi, J. R. Davies, L. Gremillet,
J. J. Honrubia, T. Johzaki, R. J. Kingham, M. Sherlock, and A. A. Solodov, Nucl. Fusion {\bf 54}, 054003 (2014).

\bibitem{Macchi13} A. Macchi, M. Borghesi, and M. Passoni, Rev. Mod. Phys. {\bf 85}, 751 (2013).

\bibitem{Yang10} X. H. Yang, Y. Y. Ma, F. Q. Shao, H. Xu, M. Y. Yu, Y. Q. Gu, T. P. Yu, Y.
Yin, C. L. Tian, and S. Kawata, Laser Part. Beams {\bf 28}, 319 (2010).

\bibitem{Jin11} Z. Jin, Z. L. Chen, H. B. Zhuo, A. Kon, M. Nakatsutsumi, H. B. Wang, B. H. Zhang, Y. Q. Gu, Y. C. Wu, B. Zhu, L. Wang, M.Y. Yu, Z. M. Sheng, and R. Kodama, Phys. Rev. Lett. {\bf 107}, 265003 (2011).

\bibitem{Green08} J. S. Green, V. M. Ovchinnikov, R. G. Evans, K. U. Akli, H. Azechi, F. N. Beg, C. Bellei, R. R. Freeman, H. Habara, R. Heathcote, M. H. Key, J. A. King, K. L. Lancaster, N. C. Lopes, T. Ma, A. J. MacKinnon, K. Markey, A. McPhee, Z. Najmudin, P. Nilson, R. Onofrei, R. Stephens, K. Takeda, K. A. Tanaka, W. Theobald, T. Tanimoto, J.Waugh, L. Van Woerkom, N. C.Woolsey, M. Zepf, J. R. Davies, and P. A. Norreys, Phys. Rev. Lett. {\bf 100}, 015003 (2008).

\bibitem{Lancaster07} K. L. Lancaster, J. S. Green, D. S. Hey, K. U. Akli, J. R. Davies, R. J. Clarke, R. R. Freeman, H. Habara, M. H. Key, R. Kodama, K. Krushelnick, C. D. Murphy, M. Nakatsutsumi, P. Simpson, R. Stephens, C. Stoeckl, T. Yabuuchi, M. Zepf, and P. A.
Norreys, Phys. Rev. Lett. {\bf 98}, 125002 (2007).

\bibitem{Ovchinnikov13} V. M. Ovchinnikov, D. W. Schumacher, M. McMahon, E. A. Chowdhury,
C. D. Chen, A. Morace, and R. R. Freeman, Phys. Rev. Letts. {\bf 110}, 065007 (2013).

\bibitem{Moore95} C. I. Moore, J. P. Knauer, and D. D. Meyerhofer, Phys. Rev. Letts. {\bf 74}, 2439 (1995).

\bibitem{Adam06} J. C. Adam, A. H\'{e}ron, and G. Laval, Phys. Rev. Letts. {\bf 97}, 205006 (2006).

\bibitem{Debayle10} A. Debayle, J. J. Honrubia, E. d¡¯Humi¨¨res, and V. T. Tikhonchuk, Phys. Rev. E {\bf 82}, 036405 (2010).

\bibitem{Davidson72} R. C. Davidson, D. A. Hammer, I. Haber, and C. E. Wagner, Phys. Fluids {\bf 15}, 317 (1972).

\bibitem{Yang94} T. Y. B. Yang, J. Arons, and A. B. Langdon, Phys. Plasmas {\bf 1}, 3059 (1994).

\bibitem{Okada07} T. Okada and K. Ogawa, Phys. Plasmas {\bf 14}, 072702 (2007).

\bibitem{Perez13} F. P\'{e}rez, A. J. Kemp, L. Divol, C. D. Chen, and P. K. Patel, Phys. Rev. Letts. {\bf 111}, 245001 (2013).

\bibitem{Molvig75} K. Molvig, Phys. Rev. Letts. {\bf 35}, 1504 (1975).

\bibitem{Karmakar08} A. Karmakar, N. Kumar, G. Shvets, O. Polomarov, and A. Pukhov, Phys. Rev. Letts. {\bf 101}, 255001 (2008).

\bibitem{Gremillet02} L. Gremillet, G. Bonnaud, and F. Amiranoff, Phys. Plasmas {\bf 9}, 941 (2002).

\bibitem{Kapetanakos74} C. A. Kapetanakos, Appl. Phys. Lett. {\bf 25}, 484 (1974).

\bibitem{Hao09} B. Hao, Z. M. Sheng, C. Ren, and J. Zhang, Phys. Rev. E {\bf 79}, 046409 (2009).

\bibitem{Volpe13} L. Volpe, D. Batani, A. Morace, and J. J. Santos, Phys. Plasmas {\bf 20}, 013104 (2013).

\bibitem{Yang15P} X. H. Yang, H. B. Zhuo, Y. Y. Ma, H. Xu, T. P. Yu, D. B. Zou, Z. Y. Ge, B. B. Xu, Q. J. Zhu, F. Q. Shao, and M. Borghesi, Plasma Phys. Control. Fusion {\bf 57}, 025011 (2015).

\bibitem{Schmitz12} H. Schmitz, R. Lloyd, and R. G. Evans, Plasma Phys. Control. Fusion {\bf 54}, 085016 (2012).

 \bibitem{Heron15} A. H\'{e}ron and J. C. Adam, Phys. Plasmas {\bf 22}, 072306 (2015).

\bibitem{Arber15} T. D. Arber, K. Bennett, C. S. Brady, A. Lawrence-Douglas, M. G. Ramsay, N. J. Sircombe, P. Gillies, R. G. Evans, H. Schmitz, A. R. Bell and
C. P. Ridgers, Plasma Phys. Control. Fusion {\bf 57}, 113001 (2015).

\bibitem{Bell93} A. R. Bell, F. N. Beg, Z. Chang, A. E. Dangor, C. N. Danson, C. B. Edwards, A. P. Fews, M. H. R. Hutchinson, S. Luan, P. Lee, P. A. Norreys, R. A. Smith, P. F. Taday, and F. Zhou, Phys. Rev. E {\bf 48}, 2087 (1993).

\bibitem{Sentoku08} Y. Sentoku and A. J. Kemp, J. Comput. Phys. {\bf 227}, 6846 (2008).

\bibitem{Gibbon05}P. Gibbon, Short Pulse Laser Interaction with Matter: An
Introduction, (Imperial College Press, London, 2005).

\bibitem{Sentoku00} Y. Sentoku, K. Mima, S. Kojima, H. Ruhl, Phys. Plasmas {\bf 7}, 689 (2000).

\bibitem{Bret10} A. Bret, L. Gremillet, and M. Dieckmann, Phys. Plasmas {\bf 17}, 120501 (2010).

\bibitem{Gremillet07} L. Gremillet, D. B\'{e}nisti, E. Lefebvre, and A. Bret, Phys. Plasmas {\bf 14}, 040704 (2007).

\bibitem{Hao12} B. Hao, W. J. Ding, Z. M. Sheng, C. Ren, X. Kong, J. Mu, and J. Zhang, Phys. Plasmas {\bf 19}, 072709 (2012).

\bibitem{Fiore10} M. Fiore, F. i\'{u}za, M. Marti, R. A. Fonseca, and L. O. Silva, J. Plasma Physics {\bf 76}, 813 (2010).

\bibitem{Silva02} L. O. Silva, R. A. Fonseca, J. W. Tonge, W. B. Mori, and J. M. Dawson, Phys. Plasmas {\bf 9}, 2458 (2002).

\bibitem{Kruer85} W. L. Kruer and K. Estabrook, Phys. Fluids {\bf 28}, 430 (1985).

\bibitem{Wilks92} S. C. Wilks, W. L. Kruer, M. Tabak, and A. B. Langdon, Phys. Rev. Lett. {\bf 69}, 1383 (1992).

\bibitem{Haines09} M. G. Haines, M. S. Wei, F. N. Beg, and R. B. Stephens, Phys. Rev. Lett. {\bf 102}, 045008 (2009).

\bibitem{Cui13} Y. Q. Cui, W. M. Wang, Z. M. Sheng, Y. T. Li, and J. Zhang, Plasma Phys. Control. Fusion {\bf 55}, 085008 (2013).

\bibitem{Chen04} Z. L. Chen, J. Zhang, T. J. Liang, H. Teng, Q. L. Dong, Y. T. Li, J. Zhang,
Z. M. Sheng, L. Z. Zhao, and X. W. Tang, J. Phys. B: At. Mol. Opt. Phys. {\bf 37}, 539 (2004).

\bibitem{Ma12} T. C. Ma, X. W. Hu, and Y. H. Chen, \textit{Principle of Plasma Physics}, (University of Science and Technology of China Press, Hefei, China, 2012).


\end{thebibliography}
\end{document}